# Tether-Based Architecture for Solar-Powered Orbital AI Data Centers

Igor Bargatin[1], Dengge Jin[2], Zaini Alansari[3], and Jordan R. Raney[4]

*Department of Mechanical Engineering and Applied Mechanics, University of Pennsylvania, Philadelphia, PA, 19096*

We propose a tether-based structural architecture for orbital data centers operating in Dawn-Dusk Sun-Synchronous (DDSS) orbits under continuous sunlight. These space-based data centers, powered solely by solar energy, could provide multi-megawatt computing for artificial intelligence (AI) inference with minimal latency to Earth. The proposed design uses a tethered chain of computing nodes with photovoltaic panels to achieve uninterrupted ~2–20 MW of computing power, and employs radiative cooling and integrated shielding to manage heat and radiation. We detail the system architecture, including mass budgets, passive attitude control, and the dynamics induced by micrometeoroid collisions.

## I. Nomenclature

*AI – Artificial Intelligence*
*CFRP – carbon-fiber reinforced polymer composite*
*CPU – central processing unit*
*DDSS – dawn-dusk sun-synchronous (orbit)*
*GEO – geostationary Earth orbit*
*GPU – graphics processing unit*
*LEO – low Earth orbit*
*MMOD – micrometeoroids and orbital debris*
*PV – photovoltaic*
*SEU – single-event upset*
*SPENVIS – Space Environment Information System*
*TPU – tensor processing unit*

## II. Introduction

Global data center energy consumption power is currently around 52 GW and is expected to double by 2026 due to increasing artificial intelligence (AI) workload demands [1,2]. Most of this growth comes from AI inference—applying trained models—rather than training new models. The concept of an orbital data center system, powered by the Sun, offers a potential solution to meet these rising demands sustainably. We propose to deploy solar-powered data centers in dawn-dusk sun-synchronous (DDSS) low Earth orbits that provide continuous sunlight for uninterrupted power [3]. By remaining in sunlight nearly 100% of the time, a DDSS orbit at approximately 1600 km altitude and the corresponding 102.5° inclination eliminates the need for energy storage and avoids thermal cycling [4]. Although other concepts for space-based data centers have been proposed [5,6,7], our approach introduces a distinct structural architecture with many potential advantages. It features a lightweight, tethered design that passively maintains optimal solar panel orientation for uninterrupted power generation, while also enhancing resilience to micrometeoroid impacts through distributed redundancy and passive angular momentum dissipation.

Each orbital data center consists of a tethered chain of computing nodes, each consisting of GPUs and CPUs as well as photovoltaic panels and a radiator for heat rejection (Fig. 1). We consider two variants: a smaller ~2 MW system deployable via medium-lift launchers like Falcon 9, and a larger ~20 MW version suited for heavy-lift vehicles such as Starship. Both offer computing power on par with a typical ~10 MW terrestrial data center, but operate independently of Earth's power grid. Thousands of such orbital data centers could collectively form a belt of artificial intelligence data centers encircling the Earth, delivering cloud computing and AI assistance to users worldwide. Our focus is on AI inference, not training, as training demands massive datasets, extremely low latency, and ultra-high-bandwidth links that are currently impractical in orbit. In contrast, inference involves single-node computation, and model parameters (~10 GB) can be preloaded onto nodes before launch and infrequently upgraded later.

---

[1] Associate Professor, Mechanical Engineering and Applied Mechanics, AIAA Member.
[2] Ph.D. Student, Mechanical Engineering and Applied Mechanics.
[3] Ph.D. Student, Mechanical Engineering and Applied Mechanics.
[4] Associate Professor, Mechanical Engineering and Applied Mechanics.

The space-based approach has several inherent advantages. It harnesses solar power 24x7 and uses it in situ, obviating the need for new terrestrial power plants or transmission infrastructure to support data center growth. It also isolates the computing infrastructure from natural disasters or grid outages on Earth, enhancing the reliability and resilience of AI services. Crucially, performing AI inference in orbit avoids the inefficiencies and safety concerns of traditional space-solar-power systems: transmitting data to and from orbit is far safer and more energy-efficient than beaming megawatts of power to Earth and routing it to ground-based data centers.

In the following sections, we describe the design of the tethered-chain orbital data center system, with a focus on structure, attitude control, and vibrations induced by micrometeoroid impacts.

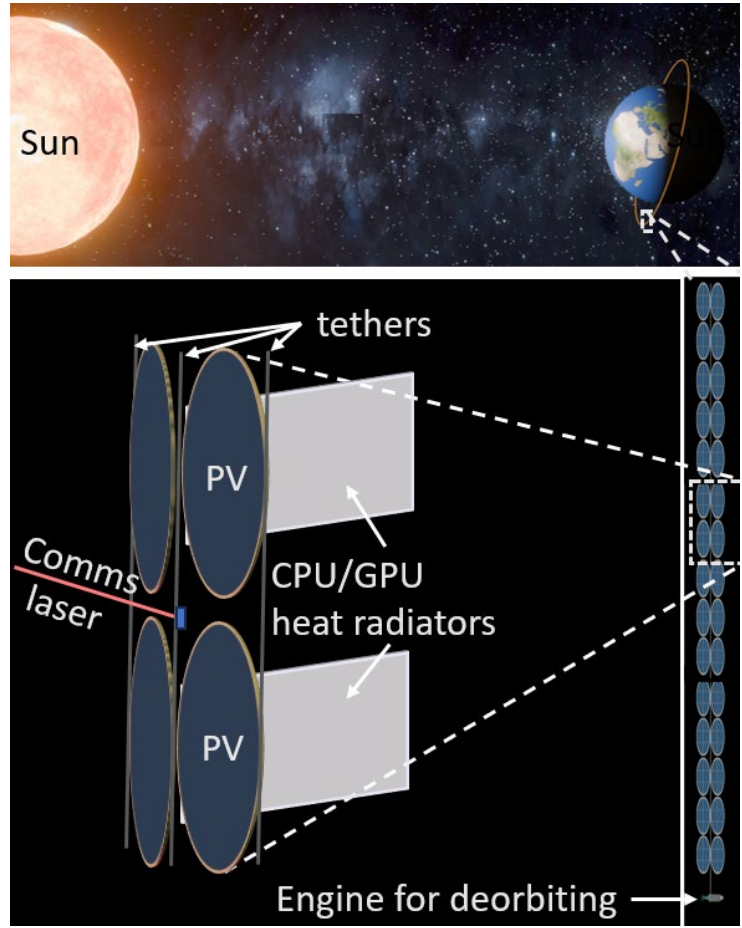

*Fig. 1:* Schematic (not to scale) of the orbital data center in a dawn-dusk sun-synchronous orbit, continuously illuminated without Earth eclipses. Bottom right: tethered-chain architecture; bottom left: individual node components with laser communication links.

### III. Proposed Architecture

#### A. Orbit Selection and Continuous Solar Power

The data centers can be placed in a dawn-dusk sun-synchronous (DDSS) orbit at ~1600 km altitude. This orbit maintains constant alignment with the terminator due to the precession of the orbital plane with Earth's annual motion [3,4]. As a result, satellites receive continuous sunlight, eliminating eclipse periods common in typical low Earth orbits. This ensures uninterrupted photovoltaic power and reduces thermal cycling. We select the lower bound of DDSS altitudes (~1600 km) to minimize communication latency and ionizing radiation doses while still avoiding Earth's shadow. Similar orbital regimes have been proposed for other space-based data center concepts [5,6,7].

#### B. Tethered Architecture and Node Design

As illustrated in Fig. 1, each orbital data center consists of thousands of identical nodes linked in a vertical, tethered chain aligned radially from Earth. This architecture spreads solar collectors and radiators over a large area, improving sunlight capture and heat rejection. The chain remains taut and stable due to gravity-gradient forces: Earth's gravity pulls the lower end downward, while centrifugal force from orbital motion pulls the upper end outward. Similar tethered systems up to 31.7 km long have flown in low Earth orbit, confirming the stability of this configuration [8].

The proposed tethered node chain architecture presents several distinct advantages over conventional truss-based [5,6] and swarm/constellation [7] designs: structural efficiency, attitude stabilization, as well as relative ease of deployment and orbital trajectory control. First, it significantly reduces structural mass compared to traditional truss-based systems. Trusses are often engineered to prevent buckling under compressive loads that can occur during attitude and orbit adjustments or as a result of impacts from micrometeoroids and orbital debris (MMOD). In a tethered-chain architectures, the primary structural elements are always in tension, effectively eliminating buckling as a design constraint and minimizing the mass of tether material (e.g., Zylon or CFRP).

Second, the tether aids attitude control. Once initial post-deployment oscillations dampen, the chain passively aligns with the local vertical—pointing radially from Earth's center to the spacecraft—providing stabilization for one rotational degree of freedom (pitch). In dawn-dusk sun-synchronous orbits (DDSS), the tether also stays nearly perpendicular to solar rays (Fig. 1), further aiding thermal and power management.

The nodes, however, can still rotate around the tether's axis (yaw). To counter this, we propose passive yaw stabilization using solar radiation pressure. By using the chevron configuration of the solar panels in Fig. 2, a torque can be generated that orients the photovoltaic (PV) panels towards the Sun, as detailed in Section IV. This passive stabilization method is particularly effective because the moment of inertia for yaw rotation is relatively small for this linear chain configuration, especially when contrasted with the much larger moments of inertia typical of kilometer-scale square array designs envisioned for space-based data centers [5].

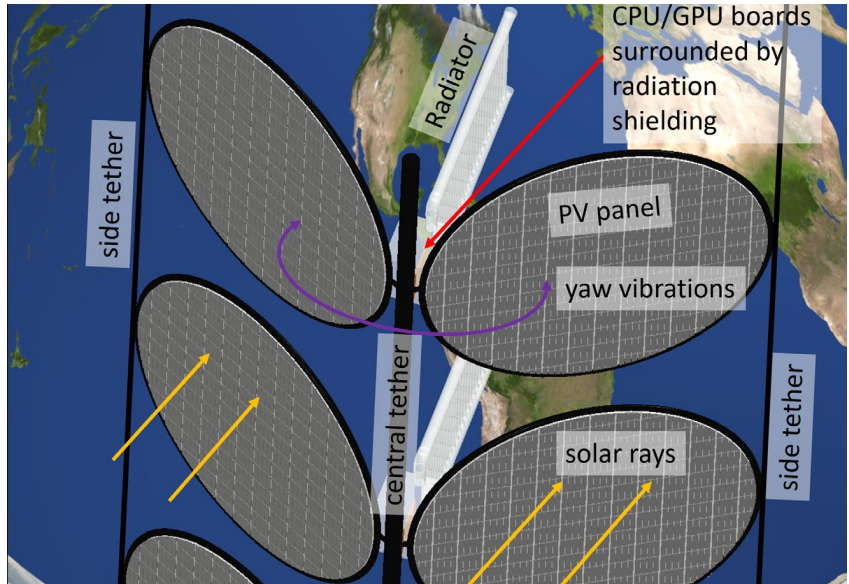

**Fig. 2:** *Details of node components. The solar panels, 3 meters in diameter each, are slightly angled into a chevron configuration to create a restoring torque when the yaw angle is different from optimal. The resulting nonlinear torsional spring operates up to yaw (rotation around the rotation axis parallel to the tethers) angles of approximately 0.5 radian (25°).*

Finally, the deployment sequence for the data center is conceptually straightforward. It can be achieved by using a high-specific-impulse thruster, such as an ion engine, to gently "tug" the chain of nodes, unreeling it from a deployment mechanism on the mothership. This contrasts with the robotic assembly of large rigid structures [5,6]. In our design, the chain length would be on the order of a few kilometers for a 2 MW system up to a few tens of kilometers for a 20 MW system, depending on the number of nodes. Each node contains computing hardware (GPU, CPU, and memory), power and thermal management systems, communications equipment, and two large photovoltaic panels for power generation. The solar panels can be ultrathin cadmium-telluride (CdTe) thin-film photovoltaics on polyimide substrates, chosen for their low weight and reported radiation hardness [9-12]. Each panel is a 3-meter diameter circular membrane less than 100 μm thick.

A 3 μm-thick CdTe cell on a 25 μm polyimide support, with a 25 μm polyimide front-side cover for protection and enhanced emissivity, weighs under 0.7 kg/m². A 25-micron-thick polyimide encapsulation provides about 60% effective thermal infrared emissivity on both sides of the PV panel, resulting in a steady-state temperature of ~360 K (90°C) under full AM0 solar flux (~1.36 kW/m²). Using even thinner protective or substrate films is problematic since the decreasing emissivity may not be sufficient to effectively reject the absorbed heat via radiation. A carbon fiber reinforced polymer hoop frame (1 cm diameter tube) can support each PV disc with minimal mass (~0.2 kg per 3 m ring). Two such discs per node, including support structure and power electronics, can have a combined mass of ~2 kg. With ~10% efficient CdTe cells (as demonstrated in prior flexible thin-film arrays [6, 7]), these panels generate approximately 2 kW of electrical power per node under full sunlight. This power sustains the node's onboard computing units (up to two CPUs and multiple GPUs with high-bandwidth memory) along with communications.

We note that a combination of tethers and photovoltaics has been previously explored in the context of bare photovoltaic tethers [13] and tether debris removal systems [14]. Such systems have relatively low PV area and do not passively orient themselves toward the Sun, reducing their power output. A gravity stabilized tether backbone was also utilized in the Suntower space-based solar power concept [15], although that system was still using many rigid structural arms for Fresnel lenses that greatly increased the moment of inertia and required careful active attitude control, likely using propellants. In contrast, our design minimizes yaw inertia by keeping all components within meters of the rotation axis, enabling passive attitude control via solar radiation pressure.

### C. Thermal Management and Radiation Shielding

Dissipating waste heat from multi-kilowatt computing nodes in vacuum is a core challenge of any orbital data center architecture. Our preliminary design uses a liquid-cooled, radiative heat rejection system that also provides radiation shielding for sensitive electronics. Each node features a closed-loop water cooling circuit—chosen for its high heat capacity and radiation shielding—circulating coolant via small pumps through cold plates on CPUs and GPUs. The coolant transfers heat to lightweight radiators. Continuous sunlight in the DDSS orbit prevents water freeze-up, maintaining thermal steady-state aided by reflected solar and Earth's infrared radiation. The water loop runs inside an aluminum heat exchanger housing that surrounds the electronics, creating a few-centimeter-thick shield against ionizing radiation and micrometeoroids [16]. Heated water flows through thin aluminum tubes with graphite fins, forming flat radiator panels. High-conductivity pyrolytic graphite sheets spread heat evenly across the radiator surface, following proven lightweight radiator designs [17].

Our preliminary thermal design achieves ~2 kW of heat rejection per node with a radiator mass of ~5 kg and area ~2 m × 3 m, maintaining electronics at about 80°C. The ~20 mm thick water+aluminum shield around each node yields an estimated ionizing radiation dose of ~10 Gray (1 krad) per year at 1600 km based on SPENVIS modeling, which is comparable to recent measurements in other orbits [18]. This ionizing radiation dose is approximately 1000 times higher than at Earth's surface, but previous NASA studies of commercial (i.e., not radiation hardened) GPUs showed that none of the GPUs experienced permanent failure at doses up to 60 Gray (6 krad) [19,20], which exceeds the dose expected over a 5-year mission. However, radiation did periodically cause functional interruptions, requiring resets and reboots. In addition, Google recently tested their Trillium tensor processing unit (TPU) under ionizing radiation and found that the high-bandwidth memory (HBM) was the most radiation sensitive part and started showing irregularities in stress tests at doses above 2 krad (20 Gy) but full machine learning workloads operated without any hard failures at doses up to 15 krad (150 Gy), which would be achieved after about 15 years on orbit [7]. These interruptions therefore should not significantly impact inference tasks, which often compute concurrently and typically complete within seconds or less, allowing fast makeup on another node in case of local failure.

### D. Communications

High-speed communication within and between the orbital data centers can be achieved using a hybrid fiber-optic and free-space laser network. Each node is connected to its neighbors via fiber optic lines running through the central tether. We propose a simple tree network topology: for roughly every 100 nodes, an optical router node aggregates traffic. Each router node contains optical switching hardware and multiple free-space laser transceivers to communicate with Earth and with other data centers. The tether's interior easily accommodates the bundle of fibers required (on the order of 100 fibers connecting to the nearest node and a smaller number of internode fibers) along with protective jacketing. Importantly, to avoid deleterious electrodynamic interactions with Earth's magnetic field, we would not use any long conductive cables along the tether [21].

For downlink and cross-link communications, the router nodes use infrared laser links. The orbital centers can beam data to existing relay satellites (such as SpaceX's Starlink constellation), which forward data to ground stations. Current Starlink satellites provide on the order of 20 Gb/s downlink capacity each and ~100 Gb/s inter-satellite laser link capacity. With over 5,000 satellites in orbit, the total available backbone bandwidth is on the order of 100 Tb/s [22]. This is well above the expected requirements of an orbital data center: even a 20 MW (10,000-node) center devoted to AI inference would need at most ~10–20 Tb/s of aggregate downlink for user data, and typical operations would be likely far less (~0.02 Tb/s per 2 MW of AI inference load) because the payload data (user queries, screen shares, video, audio, and responses) can typically be highly compressed. Thus, the existing space communication infrastructure is already close to sufficient for supporting the proposed orbital computing service, and it is rapidly improving. The internal network architecture of the data center (with router nodes forming a hierarchy) allows the entire multi-thousand-GPU array to be presented as a single compute cluster to users on Earth, simplifying resource utilization.

### E. Deployment and End-of-Life

All components of the orbital data center are designed for compact stowage during launch. The node stack can be folded and the tether coiled to fit within a rocket fairing. For example, thousands of thin disk-shaped nodes can be stacked only a few centimeters thick each, and the flexible tether can be coiled around them. Once deployed in the target orbit, the chain of nodes is gradually extended outward. One method is to use a small engine thruster at one end of the tether to pull the stack out of the container. Care must be taken to control the rate of deployment and avoid entanglement of nodes. After deployment, thrusters at both ends of the chain (leftover from the deployment system) can be used for orbit corrections, active damping of tether oscillations, and eventual de-orbiting. Compared to swarm-based data centers, which are not physically connected [7], the tethered chain data centers are much easier to deploy, perform orbital adjustments on, and de-orbit. Large swarms of independent spacecraft (e.g., millions of nodes) present significant space traffic management problems and may overwhelm the ground control's ability to prevent collisions with other satellites and debris as well as manage a huge number of independent comms links.

Since the orbital data centers will inevitably experience hardware obsolescence (state-of-the-art processors currently have a useful life of only a few years before new technology outpaces them), a plan for decommissioning is needed. At the end of a data center's ~5-year operational life, it can be safely de-orbited. The two primary options are: (1) propulsive de-orbit, using thrusters at each end of the tether to rapidly lower the orbit; and (2) solar sailing, by canting the photovoltaic panels to use solar radiation pressure to gradually reduce orbital energy. The propulsive option can de-orbit the chain in a matter of days (using chemical engines) to months (using ion engines), whereas actively changing the tilt to maximize solar radiation pressure thrust might take 1 year. In the event of a completely unresponsive data center, the high ballistic coefficient (~2 m²/kg) of the flat nodes ensures that atmospheric drag will eventually cause reentry within about 30 years even from 1600 km altitude [4].

Micrometeoroids can potentially sever the tethers connecting the nodes, which is why three of them of are necessary for redundancy (see Fig. 2). In case of catastrophic damage from a large piece of debris that severs all three tethers within the same node, the severed two parts can be deorbited using the two engines at each end of the original data center.

### F. Basic $CO_2$ footprint estimates

Launching into a 1600 km high-inclination orbit requires a delta V of approximately 11 km/s. A single 10-kg, 2-kW node launched on a partly reusable (first stage only) Falcon 9 would need approximately 650 kg of propellants (200 kg of kerosene and 400 kg of liquid oxygen), producing approximately 630 kg of direct $CO_2$ emissions during the launch. In contrast, a fully reusable future Starship would use approximately 560 kg of propellants (120 kg of liquid methane and 440 kg of liquid oxygen), producing around 330 kg of $CO_2$ emissions. In both these cases, an additional climate footprint will be associated with the production of propellants, radiative forcing of high-altitude water vapor, and any methane leaks. In contrast, operating the same node on Earth would typically emit roughly 8000 kg of $CO_2$ annually, given its yearly electricity consumption of 17,500 kWh and an average emission rate of 0.5 kg $CO_2$ per kWh from natural gas. Over a 5-year operations period, the earthbound data center will likely produce an order of magnitude more direct $CO_2$ emissions than the orbital data center.

## IV. Attitude control and vibrations of the tethered chain

By minimizing mass and yaw-axis moment of inertia, our design leverages the relatively weak solar radiation pressure for passive attitude control. Yaw oscillations can arise from deployment or sporadic micrometeoroid impacts. To analyze these dynamics, we used a discrete finite element model in ABAQUS to calculate local tension at each node (Fig. 3), then used the tension to derive the effective torsional stiffness of the three-tether configuration (one central, two side) shown in Fig. 2. Maximum tension occurs at the tether center and scales roughly quadratically with the number of nodes, making the torsional stiffness the largest in the center of the chain.

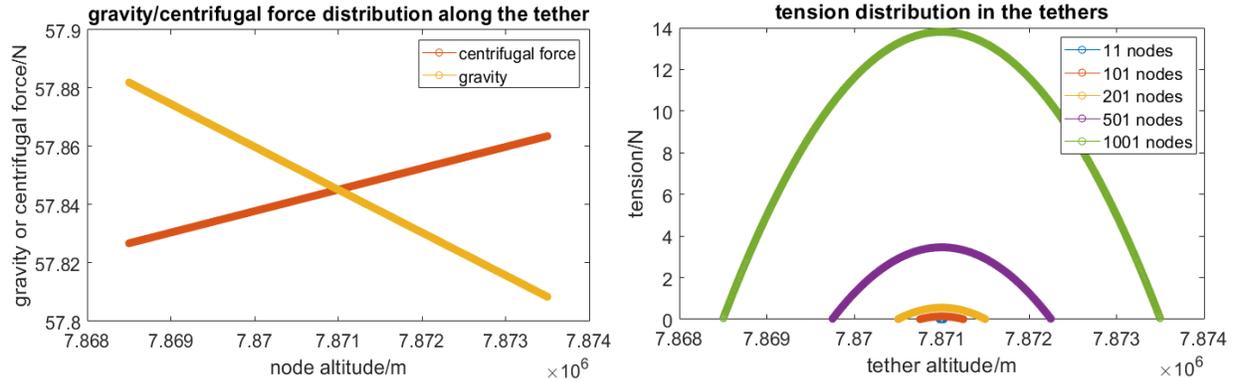

**Fig. 3:** *(left) Gravity and centrifugal forces for a distributed tethered chain architecture, with a nodal mass of 9 kg and internodal distance of 5 meters. The difference between the two forces is carried by the tension of tethers (right).*

Micrometeoroid impacts induce complex oscillations across multiple modes in the tethered chain. For our initial simulation, we modeled a 0.1-gram (~0.5 mm diameter) micrometeoroid striking the outer edge of the PV CFRP hoop—the point of maximum angular momentum transfer. The impact velocity was set at 11 km/s, perpendicular to the PV panel plane. We selected this micrometeoroid size because it is likely to embed in the hoop, transferring its full linear and angular momentum to the system. Larger particles can puncture both walls and pass through, resulting in lower momentum transfer.

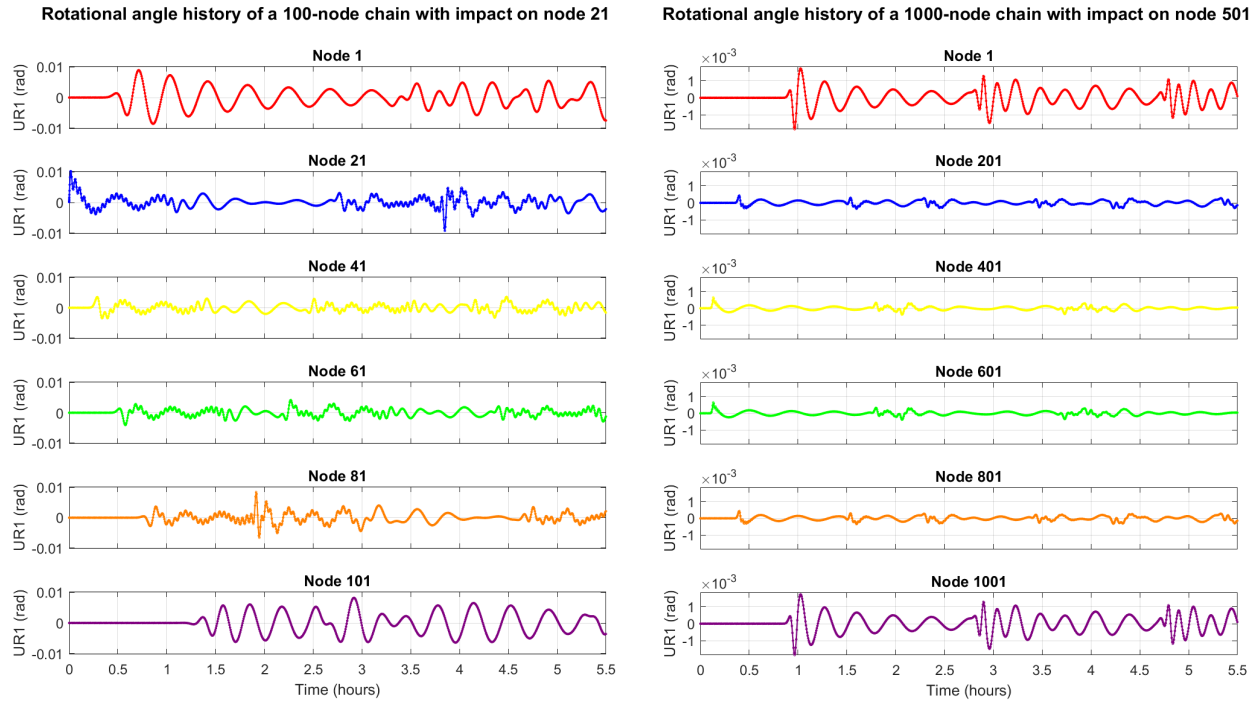

**Fig. 4:** *(left) Yaw angle oscillations for various nodes of a 101-node (left) and 1001-node (right) chain.*

The modeled oscillations are irregular, exciting multiple modes (Fig. 4). Typical periods for both pendulum-like librations and yaw rotations are less than one hour for 101- and 1001-node chains, independent of mode count. After 1–2 hours, angular momentum spreads evenly across the chain, stabilizing maximum angular deflections at roughly 0.006 radians (0.3°) for 101 nodes and 0.001 radians (0.05°) for 1001 nodes. The largest oscillations occur at the tether ends and can be easily damped using reaction wheels on the mothership and the deployment engine spacecraft. Even without active attitude control from the spacecraft at the tether ends, passive stabilization from solar radiation pressure is evident—yaw oscillations consistently center around zero (Fig. 4). This restoring torque arises from the chevron-shaped PV panels: small angular deflections alter their effective cross-sectional area to sunlight, changing the solar pressure forces and the associated torque. Unlike solar sails that maximize pressure via photon reflection, solar pressure here mainly results from light absorption [23], yet the yaw stabilization remains effective.

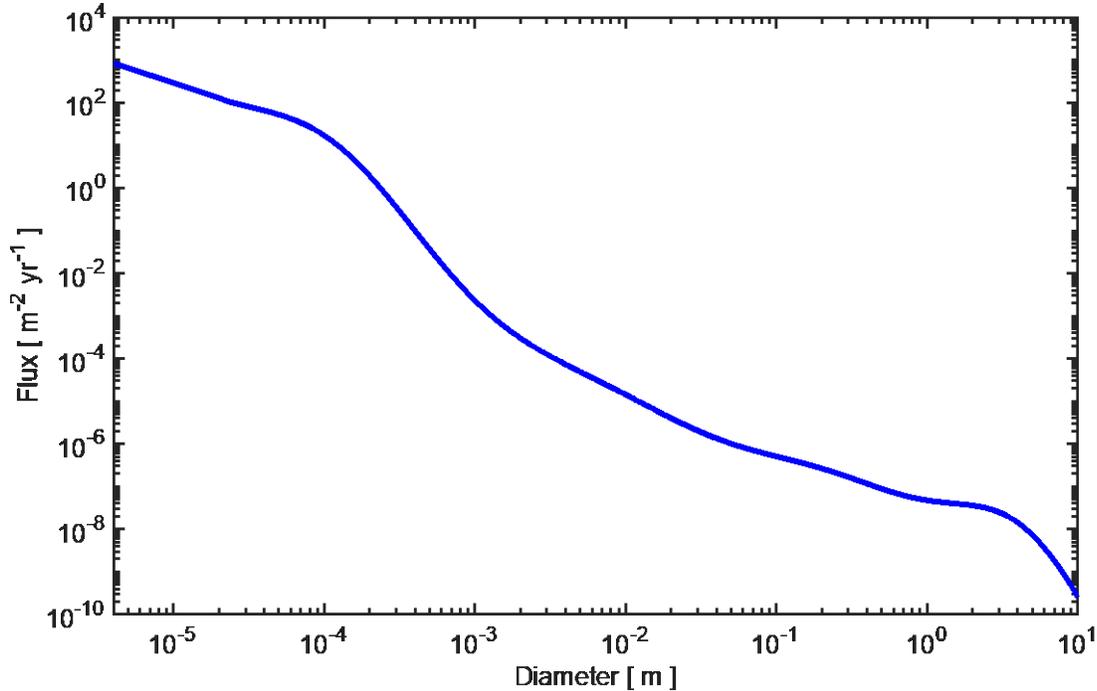

***Fig. 5:*** *Flux of micrometeoroids and orbital debris larger than a given diameter at 1600 km altitude predicted by the SPENVIS model.*

Using the SPENVIS model for a 1600-km polar orbit (Fig. 5), we also simulated how the oscillations will build over time and how the maximum yaw angles depend on the damping coefficient. Fig. 6a shows the randomly generated series of micrometeoroid impacts that match the mass-frequency distribution of Fig. 5. The location of each impact was modeled to occur randomly on the solar panels, with the micrometeoroid traveling in a randomly picked direction, resulting in realistic evolution of the yaw angles over time. The damping was introduced through a relative damping moment between neighboring nodes, $M_{damp\_i} = \beta\left(\dot{\theta}_{i+1}(t) - \dot{\theta}_i(t)\right) - \beta\left(\dot{\theta}_i(t) - \dot{\theta}_{i-1}(t)\right)$. Since these relative damping moments do not dampen out the mode where the entire chain rotates as a solid around the axis aligned with the tether, we assumed that the spacecraft at the each of the end chain maintain their orientation toward the sun fixed using reaction wheels, which are periodically unloaded using propellants.

We modeled the evolution of the average yaw angle over time for a number of different damping coefficients, as shown in Fig. 6b. After the initial buildup of oscillations, the average deflection angle stabilizes as the additional rotational energy input from micrometeoroid impacts becomes balanced by the damping. As shown in Fig. 6c, the dispersion of the yaw deflection angle, averaged over time and node number, is inversely proportional to the damping coefficient, which is the expected behavior under random perturbations. The data points show some deviation from the perfect inverse relationships, most likely because our simulation was not run for enough time to reach a steady state, especially for the lowest damping. If we extrapolate the dispersion to even smaller damping coefficients, which would be computationally expensive to simulate directly, we expect the dispersion to start reaching significant deflections of a few degrees (0.05 radian) at a critical damping coefficient of approximately $\beta_{crit} = 10^{-5}$ Nms, which is 100 times smaller than the smallest damping coefficient we were able to simulate. Given the typical moment of inertia of a few kg·m² and a typical period of the lowest modes of 2000 seconds, this damping coefficient would give the lowest modes of the chain a quality factor of approximately 50. This level of damping ($\beta_{crit} = 10^{-5}$ Nms) is therefore relatively low and can be achieved by including viscoelastic elements in the tethered chain architecture, for example, viscoelastic tethers, or viscoelastic fasteners that attach the tethers to the nodes.

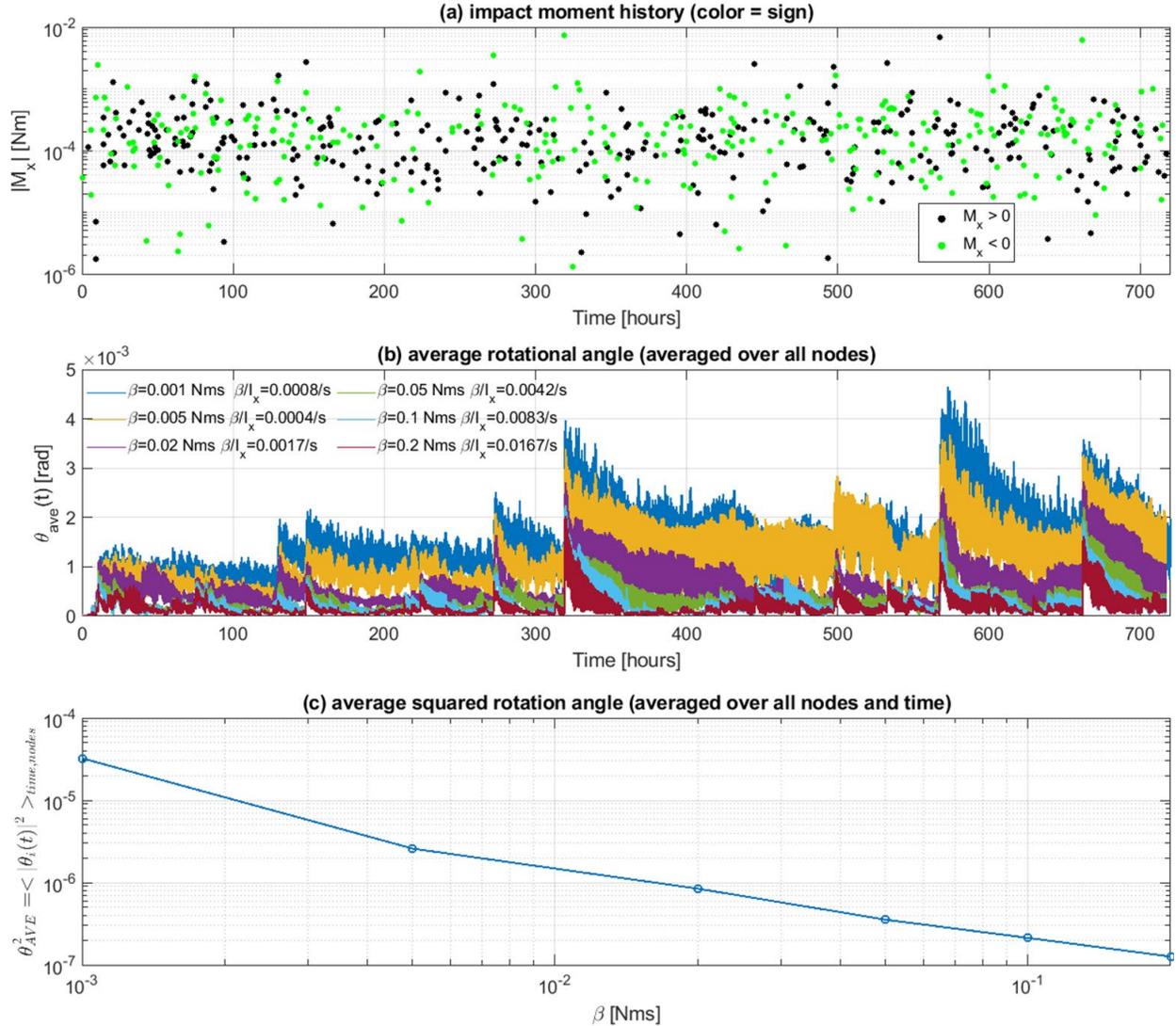

**Fig. 6:** *Dynamic response of a 100-node chain under stochastic micrometeorite and debris impacts. (a) Time history of the impact moments and the magnitude of the effective moment applied for one time step (always equal to 1 second). For each impact event, the meteorite diameter, incident direction, and impact locations on both the chain and the PV panel are sampled from their respective probability distributions determined by the micrometeorite flux. (b) Time history of the average rotational angle $\theta_{ave}(t)$. The average rotation is defined as $\theta_{ave}(t) = \frac{1}{n}\sum_{i=1}^{n}\theta_i(t)$. (c) Averaged dispersion of the rotational angle $\theta_{AVE}^2 = \langle \theta_i^2(t) \rangle_{node, time}$ as a function of the damping coefficient $\beta$.*

## V. Conclusion

We have presented a concept for orbital solar-powered data centers that could meet growing computational demands in a sustainable and resilient manner. By using continuously sunlit orbits and a distributed tethered architecture, the design provides persistent megawatt-level power and large areas for heat rejection, enabling large-scale AI inference computing in space. Our feasibility studies indicate that current-generation hardware can operate effectively in the space environment with appropriate shielding and cooling, and that existing launch vehicles and communications networks can support deployment and operation. The elimination of reliance on terrestrial power sources and the reduction in carbon emissions per compute output are key advantages. Using finite element simulations, we showed that solar pressure and modest viscoelastic damping can maintain the solar panels passively oriented toward the Sun with less than a few degrees of angular misalignment. Continued research will focus on passive attitude control, damping of micrometeoroid vibrations, and the design of lightweight heat radiators. If realized, an "Intelligence Belt" of orbital data centers could become a vital extension of Earth's technological ecosystem—one that addresses both the surging appetite for computation and the imperative for sustainability [24].

Acknowledgements: we thank Prof. Firuz Aflatouni for useful discussions. This work was supported by the School of Engineering and Applied Science at the University of Pennsylvania.